\documentclass[prd,twocolumn,eqsecnum,nofootinbib,letterpaper,superscriptaddress,showpacs,amssymb,amsmath,amsfonts,aps,altaffilletter]{revtex4}
\usepackage{color}
\usepackage{graphicx}
\usepackage{fancyhdr}
\usepackage{float}
\usepackage{ulem}
\makeatletter
\def\@fnsymbol#1{\ifcase#1\or * \or  $+$ \or  \$ \or \#  \or \dag \or \ddag \or
$\mathsection$ \or $ \mathparagraph$ \or $\|$  \or \textordfeminine \or \textbullet   \or ** \or $++$ \or  \$\$ \or \#\#  \or \dag\dag \or \ddag\ddag \or
$\mathsection\mathsection$ \or $ \mathparagraph\mathparagraph$ \or $\|\|$  \or \textordfeminine\textordfeminine \or \textbullet \textbullet \or *** \or $+++$ \or  \$\$\$ \or \#\#  \or \dag\dag \or \ddag\ddag \or
$\mathsection \mathsection\mathsection$ \or $ \mathparagraph \mathparagraph\mathparagraph$ \or $\|\|\|$  \or \textordfeminine\textordfeminine\textordfeminine \or \textbullet\textbullet\textbullet \or \else \@ctrerr\fi}
\makeatother

\newcommand\checked[1]{#1}

\newcommand\totaltime{236}
\newcommand\rhomax{15.9}
\newcommand\efficiency{0.53}

\newcommand\rpermweg{\mbox{$1.4 \times 10^{2}$}}
\newcommand\ratepermweg{\mbox{$1.7 \times 10^{2}$}}
\newcommand\nmweg{0.60}

\newcommand\dstar{51}
\newcommand\plusminus[2]{\mbox{$+$#1$/$$-$#2}}

\newcommand\textnote[2]{\textcolor{red}{\textsc{#2}}}
\renewcommand\textnote[2]{#2}

\def\thercsid{\relax}
\def\rcsid#1{\def\next##1#1{\def\thercsid{##1}}\next}
\rcsid$Id: s1short.tex,v 1.52 2003/07/28 15:41:33 patrick Exp $

\renewcommand{\today}{\number\day\space\ifcase\month\or
  January\or February\or March\or April\or May\or June\or
  July\or August\or September\or October\or November\or December\fi
  \space\number\year}

\begin{document}

\title{Analysis of LIGO data for gravitational waves from 
binary neutron stars\\
}

\date[\relax]{ RCS \thercsid; compiled \today }
\begin{abstract}
\vspace*{0.2in}

We report on a search for gravitational waves from coalescing compact
binary systems in the Milky Way and the Magellanic Clouds.  The
analysis uses data taken by two of the three LIGO interferometers
during the first LIGO science run and illustrates a method of setting
upper limits on inspiral event rates using interferometer data.  The
analysis pipeline is described with particular attention to data
selection and coincidence between the two interferometers.  We
establish an observational upper limit of
\checked{$\mathcal{R}<\ratepermweg $}~per year per Milky Way
Equivalent Galaxy (MWEG), with $90\%$ confidence, on the coalescence
rate of binary systems in which each component has a mass in the range
$1$--$3$~$M_\odot$.

\end{abstract}

%
%
%
\newcommand*{\AG}{Albert-Einstein-Institut, Max-Planck-Institut f\"ur Gravitationsphysik, D-14476 Golm, Germany}
\affiliation{\AG}
\newcommand*{\AH}{Albert-Einstein-Institut, Max-Planck-Institut f\"ur Gravitationsphysik, D-30167 Hannover, Germany}
\affiliation{\AH}
\newcommand*{\AN}{Australian National University, Canberra, 0200, Australia}
\affiliation{\AN}
\newcommand*{\CH}{California Institute of Technology, Pasadena, CA  91125, USA}
\affiliation{\CH}
\newcommand*{\DO}{California State University Dominguez Hills, Carson, CA  90747, USA}
\affiliation{\DO}
\newcommand*{\CA}{Caltech-CaRT, Pasadena, CA  91125, USA}
\affiliation{\CA}
\newcommand*{\CU}{Cardiff University, Cardiff, CF2 3YB, United Kingdom}
\affiliation{\CU}
\newcommand*{\CL}{Carleton College, Northfield, MN  55057, USA}
\affiliation{\CL}
\newcommand*{\CO}{Cornell University, Ithaca, NY  14853, USA}
\affiliation{\CO}
\newcommand*{\FN}{Fermi National Accelerator Laboratory, Batavia, IL  60510, USA}
\affiliation{\FN}
\newcommand*{\HC}{Hobart and William Smith Colleges, Geneva, NY  14456, USA}
\affiliation{\HC}
\newcommand*{\IU}{Inter-University Centre for Astronomy  and Astrophysics, Pune - 411007, India}
\affiliation{\IU}
\newcommand*{\CT}{LIGO - California Institute of Technology, Pasadena, CA  91125, USA}
\affiliation{\CT}
\newcommand*{\LM}{LIGO - Massachusetts Institute of Technology, Cambridge, MA 02139, USA}
\affiliation{\LM}
\newcommand*{\LO}{LIGO Hanford Observatory, Richland, WA  99352, USA}
\affiliation{\LO}
\newcommand*{\LV}{LIGO Livingston Observatory, Livingston, LA  70754, USA}
\affiliation{\LV}
\newcommand*{\LU}{Louisiana State University, Baton Rouge, LA  70803, USA}
\affiliation{\LU}
\newcommand*{\LE}{Louisiana Tech University, Ruston, LA  71272, USA}
\affiliation{\LE}
\newcommand*{\LL}{Loyola University, New Orleans, LA 70118, USA}
\affiliation{\LL}
\newcommand*{\MP}{Max Planck Institut f\"ur Quantenoptik, D-85748, Garching, Germany}
\affiliation{\MP}
\newcommand*{\ND}{NASA/Goddard Space Flight Center, Greenbelt, MD  20771, USA}
\affiliation{\ND}
\newcommand*{\NA}{National Astronomical Observatory of Japan, Tokyo  181-8588, Japan}
\affiliation{\NA}
\newcommand*{\NO}{Northwestern University, Evanston, IL  60208, USA}
\affiliation{\NO}
\newcommand*{\SC}{Salish Kootenai College, Pablo, MT  59855, USA}
\affiliation{\SC}
\newcommand*{\SE}{Southeastern Louisiana University, Hammond, LA  70402, USA}
\affiliation{\SE}
\newcommand*{\SA}{Stanford University, Stanford, CA  94305, USA}
\affiliation{\SA}
\newcommand*{\SR}{Syracuse University, Syracuse, NY  13244, USA}
\affiliation{\SR}
\newcommand*{\PU}{The Pennsylvania State University, University Park, PA  16802, USA}
\affiliation{\PU}
\newcommand*{\TC}{The University of Texas at Brownsville and Texas Southmost College, Brownsville, TX  78520, USA}
\affiliation{\TC}
\newcommand*{\TR}{Trinity University, San Antonio, TX  78212, USA}
\affiliation{\TR}
\newcommand*{\HU}{Universit{\"a}t Hannover, D-30167 Hannover, Germany}
\affiliation{\HU}
\newcommand*{\BB}{Universitat de les Illes Balears, E-07071 Palma de Mallorca, Spain}
\affiliation{\BB}
\newcommand*{\BR}{University of Birmingham, Birmingham, B15 2TT, United Kingdom}
\affiliation{\BR}
\newcommand*{\FA}{University of Florida, Gainsville, FL  32611, USA}
\affiliation{\FA}
\newcommand*{\GU}{University of Glasgow, Glasgow, G12 8QQ, United Kingdom}
\affiliation{\GU}
\newcommand*{\MU}{University of Michigan, Ann Arbor, MI  48109, USA}
\affiliation{\MU}
\newcommand*{\OU}{University of Oregon, Eugene, OR  97403, USA}
\affiliation{\OU}
\newcommand*{\RO}{University of Rochester, Rochester, NY  14627, USA}
\affiliation{\RO}
\newcommand*{\UW}{University of Wisconsin-Milwaukee, Milwaukee, WI  53201, USA}
\affiliation{\UW}
\newcommand*{\WU}{Washington State University, Pullman, WA 99164, USA}
\affiliation{\WU}
\author{B.~Abbott}\affiliation{\CT}
\author{R.~Abbott}\affiliation{\LV}
\author{R.~Adhikari}\affiliation{\LM}
\author{B.~Allen}\affiliation{\UW}
\author{R.~Amin}\affiliation{\FA}
\author{S.~B.~Anderson}\affiliation{\CT}
\author{W.~G.~Anderson}\affiliation{\TC}
\author{M.~Araya}\affiliation{\CT}
\author{H.~Armandula}\affiliation{\CT}
\author{F.~Asiri}\altaffiliation[Currently at ]{Stanford Linear Accelerator Center}\affiliation{\CT}
% ...
 \collaboration{The LIGO Scientific Collaboration, http://www.ligo.org}
 \noaffiliation
%
%
\maketitle
\section{Introduction}

The Laser Interferometer Gravitational-Wave Observatory (LIGO) is an
ambitious US initiative to detect gravitational waves from
astrophysical sources such as coalescing neutron stars and black
holes, spinning neutron stars, and supernovae.  
The LIGO detectors are laser interferometers with light propagating
between large suspended mirrors in two perpendicular arms.  They
measure the strain (differential fractional change in arm lengths)
produced by gravitational waves from astrophysical sources by
monitoring the relative optical phase between light paths in each
arm~\cite{Saulson:1994}. LIGO comprises three detectors housed at two
geographically distinct locations: in Hanford, WA, there are two
interferometers, one with arms $4\,\text{km}$ long (which is referred
to as H1 in this article) and one with arms $2\,\text{km}$ long (H2);
in Livingston, LA there is one interferometer with arms
$4\,\text{km}$ long (L1).
The LIGO
interferometers~\cite{Abramovici:1992ah,Barish:1999} form part of a
worldwide network of gravitational-wave detectors which includes the
British-German GEO\,600 detector~\cite{Willke:2002}, the French-Italian
VIRGO detector~\cite{Acernese:2002}, the Japanese TAMA300
detector~\cite{Tagoshi:2000bz},  and five resonant-bar
detectors~\cite{IGEC:2000}.

Among the most likely sources of gravitational waves accessible to
earth-based detectors are binary systems containing neutron stars
and/or black holes~\cite{Belczynski:2002}.  When they reach design
sensitivity, the initial interferometers in LIGO should be sensitive
to gravitational waves generated during the last several minutes 
prior to coalescence.
Current wisdom suggests that binary neutron star coalescences 
could provide up to $\sim 1/4$ events per year detectable by the initial LIGO
interferometers at design sensitivity~\cite{Kalogera:2000dz, Cutler:2001}.
Binary black hole coalescences could provide up to $\sim 2$ events
per year~\cite{Belczynski:2002}.
The rates, however, are uncertain and may be significantly lower.

Previous published
searches for gravitational waves from compact binaries used data from
the LIGO 40m prototype~\cite{Allen:1999yt} and early data from the
TAMA300 detector~\cite{Tagoshi:2000bz}.
The 40m data was taken in 1994 over a week-long
run which yielded 25 hours of data and resulted in an upper limit rate of
0.5 events per hour in the Galaxy. The instrument was sensitive to 
sources up to 25~kpc away with signal-to-noise ratio equal to 10. The
TAMA300 data was taken in 1999 over three nights which yielded 6 hours of data
and resulted in an upper limit of 0.59 events per hour for events producing a
signal-to-noise ratio larger than 7.2, corresponding to sources up to
6~kpc away.  
Searches for generic gravitational-wave bursts have also
been performed using data from multiple detectors which operated
simultaneously. Over 100 hours of data from prototype interferometers 
at Glasgow and Garching~\cite{Nicholson:1996}, and four
years of data from the International Gravitational Event Collaboration 
(IGEC) of resonant-bar detectors resulted in event rates
consistent with the background of the instrumental
noise~\cite{IGEC:2000,igec:2003}.  

This article reports on the first search for gravitational waves from
binary neutron star inspiral using LIGO data.
The first scientific data run, called S1, lasted 17 days in 2002 and 
involved all three LIGO detectors.  The detectors were sensitive
to binary inspiral events to maximum distances (at signal-to-noise 8
in a single detector) between 30 and 180 kpc, 
depending on the instrument, allowing the most sensitive search
yet.  (The TAMA300 collaboration is currently analyzing $\sim1000$
hours of
data which will provide a comparable upper limit.)
The GEO\,600 detector \cite{Willke:2002} collected data in coincidence
with LIGO during the entire S1 run and achieved an excellent duty
cycle of 98\%.  At the time of S1, GEO\,600 was still being
commissioned and was operated without signal recycling -- an essential
part of its final optical design.  It was therefore operating at a
sensitivity significantly lower than 
that of the LIGO detectors and its own target sensitivity.  Hence
GEO\,600 was not included in this analysis.
The upper limit reported here, \checked{$\mathcal{R}<\ratepermweg$}
per year per Milky Way Equivalent Galaxy (MWEG), is the best direct
observational limit on binary neutron-star coalescence to date.
This rate is far from expected astrophysical rates,  
but demonstrates the progress of instrumental commissioning and
success of the 
data analysis effort.

Many of the analysis techniques presented here
will be used in future searches for gravitational
waves. For instance, we expect to use these methods while analyzing
data taken during the second LIGO science run 
between February and
April 2003 when the detectors had roughly ten times
better amplitude sensitivity than in S1.

The outline of the paper is as follows. 
Section~\ref{s:instrument} contains a description of the instruments,
performance, sensitivity and duty cycle during S1.
Section~\ref{s:wavepop} describes in detail the target population
of binary neutron-star systems and the gravitational waves they
generate. 
The matched filtering technique used to search for these signals in
the data is reviewed in Sec.~\ref{s:analysis}. Filter outputs above
a certain signal-to-noise ratio threshold constitute triggers which
are cataloged for further analysis, provided they satisfy a $\chi^2$
test to determine the consistency of the data with the expected
waveform. 
Section~\ref{s:vetoes} describes data quality cuts and instrumental
vetoes which are applied to eliminate triggers from times when the
relevant interferometer was not operating properly.
Surviving triggers are passed through an \emph{analysis pipeline} which
generates a list of event candidates from a combination of multi- and
single-interferometer data, as detailed in Sec.~\ref{s:playground}. 
To avoid statistical bias, the veto conditions and pipeline parameters
were tuned using a \emph{playground} data set which was representative of, but
separate from, the main data set.
An upper limit on the rate of binary neutron star coalescences is
calculated in Sec.~\ref{s:results}, and systematic errors are considered in
Sec.~\ref{s:errors}.  
Section~\ref{s:future} summarizes the results and discusses the
prospects for future data runs.
\section{The LIGO Detectors}\label{s:instrument}

The LIGO interferometer design is a variant of a Michelson
interferometer, with a laser light source and a beam splitter which
directs the light along two perpendicular arms.  Mirrors at the
ends of the arms reflect the light beams back to the beam splitter,
where they recombine and interfere according to their relative optical
phase; this interference provides a sensitive measure of the length
difference between the two arms.  To augment the basic Michelson
design, partially transmitting input mirrors are placed near the
beam splitter to form a long Fabry-Perot cavity in each arm with a
finesse of $\sim 220$.  An additional partially transmitting mirror is
placed in the path of the input laser beam to form a composite
power-recycling cavity, which increases the amount of light
circulating in the interferometer.  A more detailed description of the
LIGO optical configuration and other instrumentation may be found in
Ref.~\cite{LIGOS1instpaper}.

\begin{figure*}
\includegraphics[angle=270,width=1.0\linewidth]{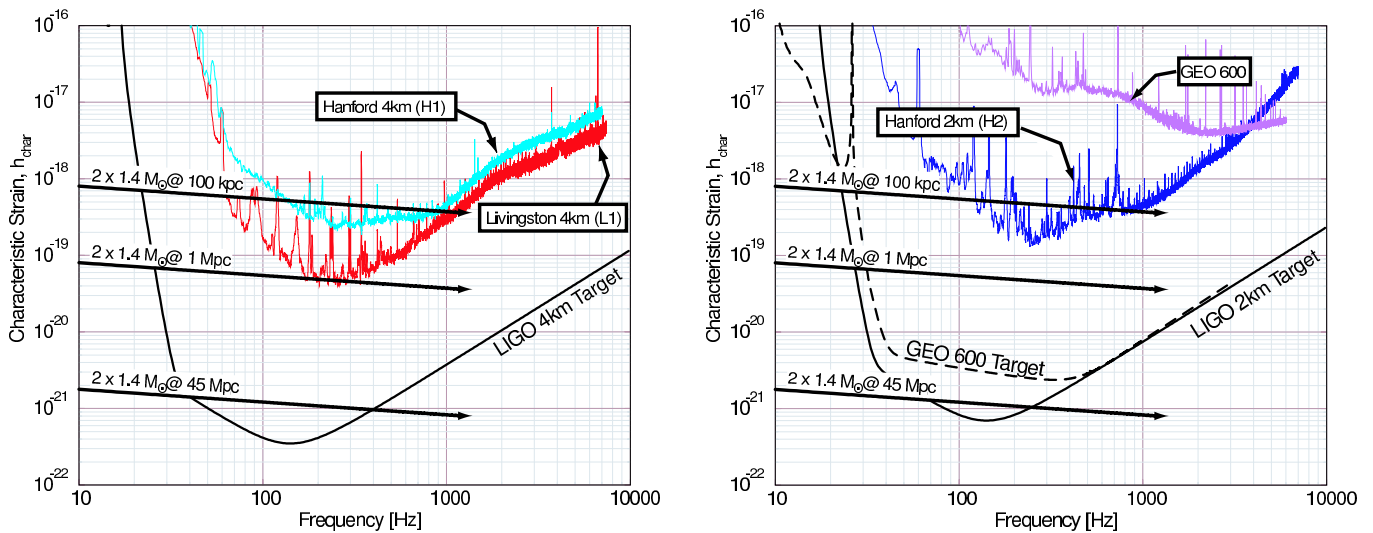}
\caption{\label{f:noisepsd}
Typical sensitivities of the LIGO and GEO\,600 interferometers during the S1 data
run, shown as equivalent RMS strain amplitude spectral density
$h_{\text{rms}}(f)=\sqrt{fS_n(f)}$, where $S_n(f)$ is the one-sided
noise power spectral density.    Typical noise spectra for the two 4km
interferometers, L1 and H1, used in our analysis are shown in the left
panel; the smooth solid curve indicates the target sensitivity of the
LIGO 4~km interferometer design.   Spectra for the 2km interferometer
H2 and GEO\,600 are shown in the right panel; the smooth solid and
dashed curves indicates the target sensitivities of the LIGO 2~km and
GEO\,600 interferometer designs.   The thick lines with arrowheads
show the characteristic strains, $h_{\text{char}}(f)=f\tilde{h}(f)$,
expected from binary neutron star systems (optimally located and
oriented with
respect to the detector) during
the last few minutes before coalescence.  These characteristic strains are
approximately equal to the amplitude of a gravitational wave signal at a given
frequency times the square-root of the number of cycles produced in a
logarithmic band about the given frequency.  The ratio of $h_{\text{char}}$ to
$h_{\text{rms}}$ in the sensitive band of the instrument provides an estimate
of the signal-to-noise ratio that could be achieved in detecting such a signal
using matched filtering.
\textnote{DHS}{
When the LIGO instruments are operating at the target
sensitivity, inspirals of double neutron stars ($ 2 \times 1.4
M_\odot$) are expected to be detectable within an equivalent volume
$\approx (4 \pi /3) \times (21  \textrm{ Mpc})^3$.
}
}
\end{figure*}

The light source for each interferometer is a medium power Nd:YAG laser,
operating at a wavelength of 1.06 $\mu$m~\cite{Savage:1998}.
Before the light is
directed into the interferometer, its frequency, amplitude and
direction are stabilized using a combination of active and
passive stabilization techniques.

To isolate the mirrors and other
elements from ground and acoustic vibrations, the detectors employ
active and passive seismic isolation systems~\cite{Giaime:1996,
Giaime:2003}, from which the mirrors are suspended as pendulums. These
form a coupled oscillator system with high isolation for frequencies
above 40 Hz. The mirrors, major optical components, vibration
isolation systems, and main optical paths are all enclosed in a high
vacuum system.  

Various feedback control systems are used to keep the
multiple optical cavities tightly on resonance \cite{Fritschel:2001}
and well aligned \cite{Fritschel:1998}.
The strain signal $s(t)=[L_x(t)-L_y(t)]/L$ is derived from the
error signal of the feedback loop used to control the differential
motion of the interferometer arms. To calibrate the error signal, the
effect of the feedback loop
gain is measured and divided out, and the response $R(f)$ to a
differential arm strain is measured and factored in. The absolute
scale of the response is established using the laser wavelength by
measuring the
mirror drive signal required to move through a given fraction of a
fringe.  The response varied over the course of the S1 run due to
drifts in the alignment of the optical elements; it was tracked
by injecting fixed-amplitude sinusoidal signals (calibration lines)
into the differential arm control loop, and monitoring the amplitudes
of these signals at the measurement (error) point
\cite{Gonzalez:2002}.

\begin{figure}
\hspace*{-0.2in}\includegraphics[height=\linewidth,angle=270]{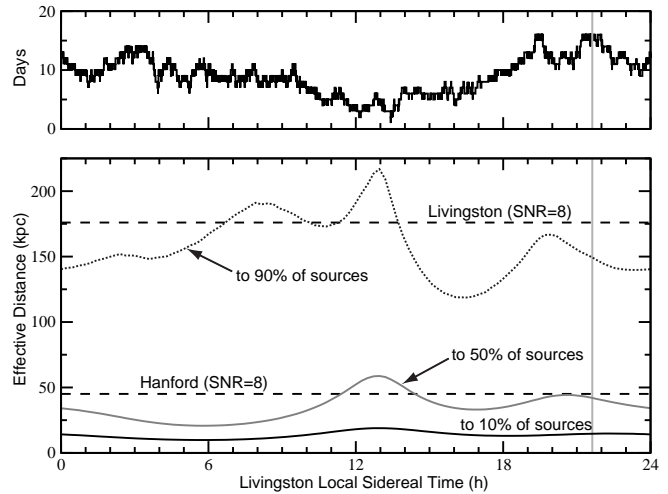}
\caption{\label{f:sensitivity}
Summary of detector status and sensitivity to the population of
neutron stars described in Sec.~\protect\ref{s:wavepop} as a function of
sidereal time.   For a given sidereal time,  the upper panel shows the
number of days during the run when at least one of the interferometers
(H1 or L1) 
was collecting scientific data.  For reference,  the vertical dotted line indicates
05:00 UTC (corresponding to midnight at Livingston) on September 01, 2002.   The lower panel shows the
effective distance as measured in Livingston [and defined by
Eq.~(\ref{e:waveform})] to 10\%, 50\%, and 90\% 
of the binary neutron 
star population described in Sec.~\ref{s:wavepop}.  The horizontal dashed lines show
the average distance at which an inspiral of $2 \times 1.4 M_\odot$ neutron
stars,  in the optimal direction and orientation with respect to each detector, 
would produce a signal-to-noise ratio of 8, {\it i.e.}\ 176~kpc
for L1 and 46~kpc for H1.}
\end{figure}

The interferometer noise is characterized by the one-sided power
spectral density $S_n(f)$ of the signal $s(t)$.  The sources of noise
that are expected to limit the eventual sensitivity of the LIGO
detectors are shot noise (determined by circulating light power,
dominant at high frequencies), thermal noise (determined by energy
dissipation mechanisms in the mirrors and suspensions, dominant at intermediate
frequencies), and seismic noise (dominant at low frequencies).
Figure~\ref{f:noisepsd} shows the expected noise due to these effects
(at LIGO's design target),
expressed as RMS strain noise, along with typical spectra achieved
by the LIGO interferometers during the S1 run.  (Typical GEO~600 noise
during S1 is also shown for comparison.)
The differences among the three LIGO spectra reflect
differences in the operating parameters and hardware implementations
of the three instruments which are in various stages of reaching the
final design configuration. For example, all interferometers
operated during S1 at a substantially lower effective laser power
level than the eventual level of 6~W at the interferometer
input. Thus the shot-noise region of the spectrum, above 200~Hz, is
much higher than the design goal. 
In addition, the S1 configuration only had a partial implementation
of the laser frequency and amplitude stabilization systems, and a
partial implementation of alignment control systems for the mirrors
and the beam splitters.  
Despite these shortcomings, the
detectors were sensitive to binary neutron star coalescences within
the Galaxy and the Magellanic Clouds as illustrated in
Fig.~\ref{f:sensitivity}.

The 17-day run yielded 363 hours of data
when at least one interferometer was in stable operation.  The three
interferometers were simultaneously in stable operation for 96 hours.
For the analysis presented in this article, we chose to use data only
from the two 4~km detectors, L1 and H1.  While H2 was nearly as sensitive as
H1,  its noise exhibited a greater degree of non-stationarity,
leading to a rate of spurious triggers which would have compromised
the sensitivity of the search.
L1 and H1 were simultaneously
operational for 116 hours during the S1 run, providing data for the
first combined analysis of interferometric detectors sensitive to
inspiral events throughout the Galaxy.  In addition, they were separately
operational for 54 and 119 hours, respectively.

\section{Target population and Waveforms}\label{s:wavepop}
Radio observations of pulsars confirm the existence of binary neutron
star systems in the Galaxy~\cite{Hulse:1994,Taylor:1994}.   General
relativity predicts the decay of a binary orbit due to the emission
of gravitational radiation. The decay rate inferred from observations
of PSR1913+16 agrees with the prediction within
0.3\%~\cite{Taylor:1982,Taylor:1989,Weisberg:2002}.  The orbital decay
is easily modeled for compact binary systems containing neutron stars or 
stellar mass black holes.  The binary orbit is expected to evolve through
the LIGO frequency band
by the emission of gravitational waves alone, making it possible to
accurately compute the evolution without reference to complicated
micro-physics. 

When a compact binary system first forms,  the orbit may be widely
separated and highly eccentric. (See Ref.~\cite{Belczynski:2002} for a discussion
and plots of birth separations and eccentricities).   Gravitational radiation,
emitted 
predominantly at twice the orbital frequency of the binary system,
causes the orbit to shrink and circularize (much faster than it
shrinks~\cite{Peters:1964}) so that the binary components eventually
spiral together along a sequence of nearly circular orbits with
decreasing period.  For binary neutron
stars or stellar-mass black holes, the gravitational radiation
eventually enters the frequency band of earth-based gravitational-wave
detectors.  At this point, the orbit decays rapidly and the
gravitational waveform chirps upward in frequency and amplitude,
sweeping through LIGO's sensitive band.
During S1,  the LIGO interferometers were
sensitive to gravitational-wave frequencies above about 100~Hz;
an inspiral signal from two $1.4M_\odot$ objects would traverse the sensitive
band in 2~seconds.    At design sensitivity, the sensitive band will
stretch down to $\simeq 40$~Hz and the signals will spend
about  30~seconds in the sensitive band.   

For low-mass binary systems,
the waveforms are well approximated by a post-Newtonian
expansion~\cite{Blanchet:1995ez,Blanchet:1996pi,Damour:2000zb} in the
LIGO frequency band.  
Due to the uneven convergence of this expansion and a still
indeterminate coefficient at higher order,  we used second-order 
post-Newtonian waveforms~\cite{Blanchet:1996pi} in this analysis.
The waveforms are parameterized by the masses of the two companions 
$I=(m_1,m_2)$, the
inclination of the orbit relative to the plane of the sky, and the starting
orbital phase.  
Other orbital parameters such as eccentricity and spin
are not expected to be significant for binary neutron star
coalescence~\cite{Bildsten:1992my,Apostolatos:1995,Belczynski:2002}, so we do not consider them in this analysis.  The strain produced
in the instrument is written as
\begin{equation}
h(t) = \frac{1 \textrm{ Mpc}}{D_{\mathrm{eff}}} 
\bigl[ \sin \alpha \, h^I_s(t-t_c) + \cos\alpha \, h^I_c(t-t_c) ] \; ,
\label{e:waveform}
\end{equation}
where $\alpha$ depends on the orbital phase and orientation of the
binary system,  $t_c$ is the time (at the detector) when the binary reaches its
inner-most stable circular orbit,  and $h^I_{s,c}(t-t_c)$ are the two
polarizations of the gravitational waveform produced by an inspiralling
binary that is optimally oriented at a distance of 1 Mpc.   An optimally-oriented
binary system is one that lies on the detector's $z$-axis with its
orbital plane parallel to the $x$-$y$ plane, defined by the
arms of the detector (i.e., directly above or below the detector and
orbiting on the plane of the sky).    The effective distance
$D_{\mathrm{eff}}$ depends on the true distance $r$ to the binary, its
location in the sky relative to the detector, and its orientation.
This dependence is, in part, caused by the non-uniform detector response over the sky.
If the source is not optimally oriented, then $D_{\mathrm{eff}}>r$.
The binary
inspiral waveform can thus be parameterized (for a single detector) in
terms of the component masses, the effective distance, and the signal
phase.

The rate at which neutron star binaries coalesce in our Galaxy can be
estimated using the observed sample of binary pulsars. (See,  for
example,  Ref.~\cite{Kalogera:2000dz}.)  This rate estimate can be
extrapolated to extra-galactic distances (following Phinney~\cite{Phinney:1991ei}) by assuming that the
coalescence rate is proportional to the formation rate of massive
stars and that the primordial binary population in our Galaxy is
typical.  Since the rate of massive star formation is proportional to
blue-light (B-band) luminosity,  the number of coalescences
contributed by another galaxy is determined by the ratio of its
blue-light luminosity to that of the Milky Way.  The sample population
for our analysis used spatial and
mass distributions from a Milky Way population produced by the
simulations of Ref.~\cite{Belczynski:2002} with the spatial
distribution described in Ref.~\cite{Kim:2002uw}.  
Additional sources from the Large
and Small Magellanic Clouds, treated as points\footnote{The
angular diameters of the Large and Small Magellanic Clouds are $7$
and $4$ degrees, respectively.  These are comparable to
the best angular resolution
that can be achieved in our analysis using time of arrival information from two
LIGO detectors to determine sky position information.  The resolved
variations of instrumental response across the Magellanic Clouds is
negligible in our analysis.
}
at their
known distances and sky positions, were also added.   The number of
sources was
proportional to the absolute blue-light luminosity of the LMC and SMC,
with correction
factors applied to account for reddening and the lower metallicity of
these objects. The latter leads to lower neutron star formation rates
primarily due to weaker stellar winds, which in turn favor the formation
of more massive compact objects.  With these corrections, the event rates
from the Large and Small Magellanic Clouds are taken to be 11\% and 2\%
of the Milky Way rate.  
We note that this population model may not be exactly accurate, but is
representative of the current understanding of binary neutron star
formation.

\section{Template Based Trigger Generation}\label{s:analysis}
The data stream from each detector was searched for inspiral
waveforms using matched filtering, {\it i.e.}, by evaluating the
correlation (with a frequency-dependent weighting to suppress noise) between
the data and a template waveform for all possible coalescence times.
We use templates for non-spinning binaries, so each waveform is identified
by a mass pair $I=(m_1,m_2)$,  a phase $\alpha$ and a distance
$D_{\mathrm{eff}}$ as described above.   The gravitational wave
signals also obey the approximate relationship
\begin{equation}
\tilde{h}^I_c(f) = -i \tilde{h}^I_s(f) \; ,
\label{e:symmetry}
\end{equation} 
where $f>0$ and the Fourier transform $\tilde{q}(f)$ is defined by
\begin{equation}
\tilde{q}(f) = \int_{-\infty}^{\infty} e^{-2 \pi i f t} q(t) \, dt \;
.
\end{equation}
We exploit the symmetry (\ref{e:symmetry}),  which is exact within the
stationary-phase approximation used in this analysis,\footnote{The 
stationary-phase approximation to the Fourier transform of inspiral 
template waveforms was shown to be sufficiently accurate for gravitational-wave 
detection in Ref.~\cite{Droz:1999qx}.} to reduce computational 
overhead in searching over the phase $\alpha$.   If the detector's calibrated
strain data is $s(t) = n(t)+h(t)$, where $n(t)$ is the instrumental
strain noise and $h(t)$ is a gravitational wave signal (if present),  then
the matched filter output for given masses $I=(m_1,m_2)$ is the
complex time series
\begin{equation}
\label{e:xfilter}
  z(t) = x(t) + iy(t)= 4\int_{0}^{\infty}
    \frac{\tilde{h}^I_c(f)\tilde{s}^\ast(f)}{S_n(f)}\,e^{2\pi ift} df
\end{equation}
where $S_n(f)$ is the one-sided strain noise power spectral density. 
In this expression,  $x(t)$ is the
matched filter response to the $\alpha=0$ waveform $h^I_c$ while $y(t)$ is the
matched filter response to the $\alpha=\pi/2$ waveform $h^I_s$.  
Matched filtering
theory~\cite{wainstein:1962} provides a simple way to search over the 
phase $\alpha$:  construct the signal-to-noise ratio (SNR) of the
matched filter output,
\begin{equation}
\label{e:snr}
  \rho(t)=\frac{|z(t)|}{\sigma} ,
\end{equation}
where 
\begin{equation}
\label{e:variance}
  \sigma^2 = 4\int_{0}^{\infty} \frac{|\tilde{h}^I_c(f)|^2}{S_n(f)}df 
\end{equation}
is the variance
of the matched filter output due to detector noise.
For
stationary and Gaussian noise, $\rho$ is the optimal detection statistic
for a single detector.   

The waveform (\ref{e:waveform}) depends on the masses of the two companions, 
so a \emph{bank} of
templates that covers the expected range of neutron star masses must be used
\cite{Owen:1998dk}.  We adopted a template bank that covers the mass range
\checked{$1\mbox{--}3\,M_\odot$} for each companion.  The discrete bank was
designed to cause less than \checked{$3\%$} loss in SNR
due to parameter mismatches between any waveform and the nearest template in 
the bank.  The layout of the template bank depends on the noise power
spectral density of the instrument.   A single template bank was used
in this analysis:   banks were first generated for each instrument and
the bank with the most templates (in this case, the one generated for L1)
was used.   We checked that 
the resulting 2110 templates covered the mass range with \checked{$\leq 2\%$} 
loss of SNR for L1 and \checked{$\leq 7\%$} loss for H1.  Waveforms
with total mass below $4.0 M_\odot$ incurred \checked{$\leq 3\%$} loss
of SNR in both instruments.  Using a single template bank
allows easier comparison of inspiral candidates in the coincidence step of our 
analysis.

To reject transient noise artifacts that may excite a matched filter, but do
not accumulate SNR as a chirp signal would, we employed an additional
time-frequency veto in which the contribution to the filter output $z(t)$
from $p$ frequency sub-bands is compared to the expected contribution
for the templates~\cite{Allen:1999yt,grasp}. The frequency sub-bands were
chosen so that the expected
chirp would produce an equal contribution to both the real and imaginary
components of the filter output from each sub-band. 
The chirp for each
sub-band is filtered to produce the $p$ complex-quantities $z_l(t)$
and the statistic is constructed as
\begin{equation}
  \chi^2(t) = \frac{p}{\sigma^{2}}\sum_{l=1}^p |z_l(t)-z(t)/p|^2.
\end{equation}
In the presence of Gaussian noise alone, $\chi^2$ is chi-squared
distributed with $\nu=2p-2$ degrees of freedom.   In this analysis,
we did not optimize over different values of $p$,  but chose 
$p=8$ which worked well.   

If a putative signal $h(t)$ has masses which do not exactly match any
template in the bank, then $\chi^2$ has
a non-central chi-squared distribution with $2p-2$ degrees of freedom and a
non-central parameter $\lambda=2\rho^2\varepsilon^2$, where $\rho$ is the SNR for
the signal and $\varepsilon$ is the fractional loss of SNR due to
parameter mismatch.  While it is possible to construct constant confidence
thresholds on the non-central chi-squared distribution for various signals,
in this analysis we simply require
\begin{equation}
\chi^2 < 5 (p+ 0.03 \rho^2)
\label{e:chisq-threshold}
\end{equation}
for any inspiral event, where $p=8$ as described above.  We
refer to this cut as the $\chi^2$-veto.  Since the detector noise was not
Gaussian,  the threshold was selected based on performance in the playground
data set described in Sec.~\ref{s:vetoes} and not using the exact result for the non-central chi-squared
distribution.

We identify possible inspirals in a single detector (H1 or L1) by
finding maxima of $\rho(t)$ above a
certain threshold (chosen to be $\rho^\ast = 6.5$ in this analysis),
subject to the $\chi^2$-veto constraint of Eq.~(\ref{e:chisq-threshold}), and
separated in time by at least the length of the template.  Each such
maximum is considered a \emph{trigger};  the inferred coalescence
time, $\rho$, and $\chi^2$ values are cataloged in a database along with the
template parameters and effective distance (in Mpc),
$D_{\text{eff}}=\sigma/\rho$.

\begin{figure}
\includegraphics[width=\linewidth,angle=0]{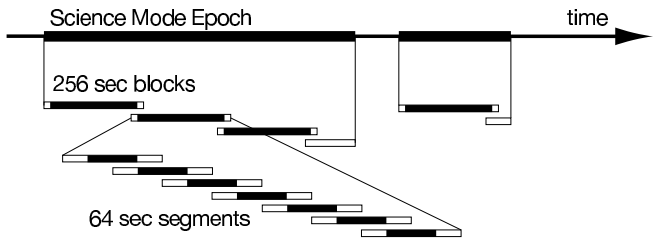}
\caption{\label{f:segments}
Times when an interferometer was in stable operation were identified as
science mode epochs indicated by the thick black lines at the top of
the figure.   These
science mode epochs were analyzed in blocks of 256 seconds overlapped
by 32 seconds (indicated in white).  If there was not enough data at the end of a science
mode epoch to take a 256 second block for analysis, the extra data was
dropped from the analysis.   Each of these blocks were further divided
into 7 overlapping segments of 64 seconds which were 
then searched for inspiral signals.  
The overlaps are needed to avoid contamination in the correlation used
to compute the SNR.}
\end{figure}
Times when each interferometer was in stable operation were identified as
science mode epochs.  These science mode epochs were analyzed in
blocks of 256 seconds overlapped by 32 seconds as shown in
Fig.~\ref{f:segments}.  If there was not
enough data at the end of a science mode epoch to take a 256 second
block for analysis, the extra data was dropped from the analysis.
Each 256 second block was read by the LIGO Data Analysis System
(\textsc{ldas})~\cite{LDAS}, which down-sampled it from 16~kHz to 4~kHz.
The power spectrum of the data was estimated for each block by
dividing it into four 64 second segments and taking the mean power
spectrum of these four segments.  The matched filter given in
Eq.~(\ref{e:xfilter}) was implemented on 64 second data segments using
routines in the LSC Algorithm Library (\textsc{lal})~\cite{LAL}.%
\footnote{The analysis was performed on the \textsc{medusa}
computing cluster at the University of Wisconsin--Milwaukee
\texttt{http://www.lsc-group.phys.uwm.edu/beowulf/medusa}}  In order
to avoid end-effects in performing the correlation described by
Eq.~(\ref{e:xfilter}), we modified $1/S_n(f)$ so that its inverse
Fourier transform had a maximum duration of $\pm$16 seconds. 
The first and last 16 seconds of each filtered 64
second segment were ignored as corrupted by the end-effects of the
filter.  The 64 second segments were overlapped by 32 seconds---thus
forming 7 overlapping segments in each 256 second block---so that no
data was lost within each block.  Since the blocks were also
overlapped by 32 seconds, only the first 16 seconds of data from the
first block and the last 16 seconds of data from the last block were
lost from each science-mode epoch.    These effects combined result in
the loss of 14 hours of data from each of the L1 and H1
interferometers.

When the interferometers at Hanford and Livingston were in stable
operation,  we checked for coincident signals to
improve confidence in a detection.
Since the Hanford and Livingston detectors are approximately
co-aligned, they should observe essentially the same
gravitational-wave signal.\footnote{%
The two LIGO interferometers H1
and L1 are not exactly aligned due to the curvature of the earth.  The
effect of this curvature is to introduce small differences in response
of each instrument to a real gravitational wave.  We have ignored this
effect at the present time, but plan to include it in future analyses.}
Ignoring mis-alignment and assuming the instrumental noise
is Gaussian and uncorrelated, the optimal detection statistic can be
written as
\begin{equation}
  \label{e:coherentsnr}
  \rho^2_{\text{coherent}}(t) = \max_\tau \frac{|z_{\text{L1}}(t)+z_{\text{H1}}(t+\tau)|^2}{\sigma_{\text{L1}}^2+\sigma_{\text{H1}}^2}
\end{equation}
where $z_{\text{L1}}(t)$ and $z_{\text{H1}}(t)$ are the complex matched filter
outputs from the L1 and H1 detectors, $\sigma_{\text{L1}}^2$ and $\sigma_{\text{H1}}^2$ are the variances of
these matched
filter outputs for the two detectors, $\tau$ is the difference in the arrival
time of the signal between the two detectors, and the maximization is performed
over all possible values of $\tau$ up to the light-travel time between the two
detectors ($\pm10$~ms)~\cite{Pai:2000zt,Finn:2000hj}.  This statistic uses the same
template in each instrument and assumes that the time of arrival is
consistent with the light travel time between the instruments.    
Since $\sigma$ [Eq. (\ref{e:variance})] depends
on the inverse power spectral density, a large value indicates good
sensitivity.  If, for example, L1 is considerably more sensitive than
H1 (as it was during S1), then  $\sigma_{\text{L1}}\gg\sigma_{\text{H1}}$.
Thus, one has $|z_{\text{L1}}|\gg|z_{\text{H1}}|$ both during typical
operation and when a signal is present,
and a good approximation to the coherent statistic is
\begin{equation}
  \label{e:approxsnr}
  \rho^2_{\text{coherent}}
  \simeq |z_{\text{L1}}|^2/\sigma_{\text{L1}}^2 = \rho_{\text{L1}}^2 \; .
\end{equation}
Since L1 was much more sensitive than H1 during the S1 run,
$\rho_{\text{coherent}}$ for an event seen while both detectors were operating
is well approximated by the $\rho$ value for L1 alone;  when only H1
was operating,  $\rho_{\text{coherent}}$ reduces to the $\rho$ value
for H1 since the contributions from L1 vanish.   
We also note that a binary inspiral signal would have $\rho_{\text{L1}} \gtrsim 4
\rho_{\text{H1}}$,  so a genuine signal would not produce a trigger
in H1 unless it appears in L1 with very high SNR (greater than $\sim 26$).

\section{Data Quality Criteria and Vetoes}\label{s:vetoes}
The performance of the LIGO interferometers varied significantly
during the S1 run on both long and short time scales.  We omitted
intervals of data from a given interferometer if it was not properly
calibrated or if it had an unusually high level of noise, as described
below.  We also were able to \emph{veto} some individual triggers
which had a clear instrumental origin.  To avoid statistical bias, the
specific veto criteria were decided based on studies of a
\emph{playground} data set comprising roughly 10\% of the data
collected when all three interferometers were operating.  This data
was excluded from calculation of the final analysis results.

\subsection{Instrumental calibration}
\label{ss:bad-cal}

As mentioned in Section~\ref{s:instrument}, the time variation of the
interferometer response was tracked by continuously injecting
sinusoidal signals with known amplitudes. 
The calibration was updated once per
minute, and the analysis of each 256-second block of data used the
first available calibration update within the block.  There
were periods of time when the sinusoidal injections were absent, however, and
the calibration could not be updated.  Blocks of data in which such a
calibration drop-out occurred were not analyzed.  There were also some
periods of time when H1 calibration information was present but was
deemed unreliable; these periods also were omitted from the analysis.
In total, 17 hours of H1 data and 8 hours of L1 data were omitted from
the analysis because of missing or unreliable calibration data.

\subsection{Noise level}
\label{ss:epoch}

The noise in the gravitational-wave channel of each interferometer was
sensitive to optical alignment, servo control settings, and
environmental conditions.  During most of the run, the noise level
varied by less than a factor of two; however, there were a number of
times when the noise level was significantly higher.  We chose to omit
these periods when the noise was particularly high.  The specific
criteria were developed by the working group searching for
gravitational-wave bursts and adopted for the inspiral analysis as
well.  Each interferometer's performance was tracked by calculating
the band-limited root-mean-square noise (BLRMS) in four frequency
bands $\{B_1, B_2, B_3, B_4\} = $\{320--400~Hz, 400--600~Hz,
600--1600~Hz, 1600--3000~Hz\}.  For each band, the noise power
$P_i(t)$ was calculated every $1/8$~seconds, then averaged over
360-second time intervals and compared to the mean value  $\bar{P}_i$ for all
science-mode data collected.  Based on empirical studies of
correlations between the power in each band and non-stationarity of
the noise, we decided to eliminate any contiguous epoch of science
data if there was any 360-second interval during the epoch for which
$P_1 > 10\,\bar{P}_1$ or $P_j > 3\,\bar{P}_j$ for $j = 2,3,4$.  This
BLRMS cut removed 13 hours (8\%) of the L1 data and 43
hours (18\%) of the H1 data.

Since the BLRMS cut uses the noise in the gravitational-wave
channel to identify times when data quality is suspect, a sufficiently
strong inspiral signal could potentially cause the veto to be invoked.
Based on the known amplitude response of the instruments,  we
determined that a binary neutron star inspiral signal would be vetoed
in this way only if it were closer than $\sim 300$~pc, corresponding
to a SNR of $4.7 \times 10^3$ in L1.  By way of confirmation,  we also
computed $P_i$ for periods when large-amplitude simulated inspiral
waveforms were injected into the interferometers.   The observed
safety margin was consistent with the model calculations.    Since
$\ll 1\%$ of the target population is within 300~pc of
Earth,  the systematic effects of the BLRMS cut on our search were 
negligible.

\subsection{Instrumental vetoes}
\label{ss:vetoes}

The data quality cuts described above addressed performance variations
over long time scales.  Each of the interferometers also exhibited
non-stationary behavior on short time scales, with occasional
glitches and/or brief periods of elevated broadband noise in the
gravitational-wave channel.  Because the matched filtering technique
used in this analysis assumed the noise spectrum to be stationary over
periods of several minutes, these transients tended to excite the
inspiral filter bank in such a way as to be recorded as triggers with
fairly large SNR, even though they did not closely resemble the
waveform of an inspiral.  The $\chi^2$~veto [Eq.~(\ref{e:chisq-threshold})]
eliminated many of these
triggers, but some remained, appearing as a high-side tail in the SNR
distribution of inspiral triggers found in the playground data set.

We attempted to identify environmental or instrumental origins for
these high-SNR triggers by checking for coincident transients
in the many auxiliary data channels which were recorded along with the
gravitational-wave channel.  These included environmental monitoring
sensors (seismometers, accelerometers, magnetometers, etc.)\ as well
as various signals related to the operation of the interferometers.
We evaluated several transient-detection algorithms, eventually
choosing a simple one which applies a high-pass filter to the data and
records excursions from zero which exceed a given size threshold.  We
developed an automated procedure to veto any inspiral trigger
within a given time window around auxiliary-channel glitches found by
this algorithm.  For each of several promising auxiliary channels, the
excursion size threshold and time window were tuned using the
playground data set to maximize the number of triggers vetoed
without introducing undue dead-time.  The results of these studies for
each interferometer are summarized below. 

The H1 detector experienced distinct glitches in the
gravitational-wave channel at a rate of about 4 per hour.  Although no
external environmental cause was identified, nearly all of
these glitches were clearly visible in an auxiliary channel derived from a
photo-diode at the interferometer's reflected port.  This channel is
sensitive to the average arm length and is used to control the
frequency of the laser light.  We vetoed inspiral triggers
within a $\pm 1$~second window on either side of glitches found in this
auxiliary channel; this veto condition introduced 
a dead-time of \checked{0.2\%}.   
Based on the detector design, a real gravitational wave would not be
expected to appear with a significant amplitude in this auxiliary
channel; we verified this experimentally by injecting simulated
inspiral waveforms into the interferometer arm length control servo (changing
the arm lengths using electromagnetic actuation to push the suspended
mirrors) and observing the signal strength in this and other auxiliary
channels.

High-SNR inspiral triggers in the L1 detector were 
strongly correlated with transients in an auxiliary channel derived
from the photo-diode at the interferometer's antisymmetric port,
nominally orthogonal in demodulation phase relative to the
gravitational-wave channel.
This auxiliary channel was not used to control any
degree of freedom in the interferometer;  it was sensitive to imbalance in the
modulation sidebands and to alignment fluctuations.
This suggested its use as a veto channel.
Unfortunately, simulated inspiral waveforms
injected into the arm length control servo appeared with non-negligible
amplitude in this auxiliary channel.  We suspect this was an artifact of
injecting a large signal with imperfectly balanced mirror actuators,
introducing an oscillatory misalignment.  To be safe, however, we chose
not to veto based on this channel.  No other auxiliary channel offered
an efficient veto, so no instrumental veto was applied for L1.

\section{Analysis Pipeline and Tuning}\label{s:playground}

\textnote{PRB}{The detection of a gravitational-wave inspiral signal
in the S1 data would (at the least)} require triggers in both
\textnote{PRB}{L1 and H1}
with consistent arrival times (separated by less than the
light travel time between the detectors) and waveform parameters.
Such a temporal coincidence requirement has the advantage of greatly
reducing the background rate due to spurious triggers in the
individual detectors.
It limits the volume of space
searched to that which can be seen by the \emph{less} sensitive
detector, however, and it limits the observation time to the periods of
simultaneous operation.  Because the L1 detector was much more
sensitive than H1 during the S1 run, and because they operated
simultaneously less than $30\%$ of the time, we developed a more
sophisticated \textnote{PRB}{(upper-limit)} 
analysis pipeline which makes use of triggers from the individual detectors 
when a coincidence test is not possible.  
Studies
of the playground data set indicated that the additional background
rate introduced by this choice should not offset the improvement in
event rate limit that comes from increased observation time.  Of
course,
event candidates identified during non-coincident observation times could
not lead to an unambiguous detection of gravitational waves. 

\begin{figure}
\begin{center}
\hspace*{-0.2in}\includegraphics[angle=270,width=0.9\linewidth]{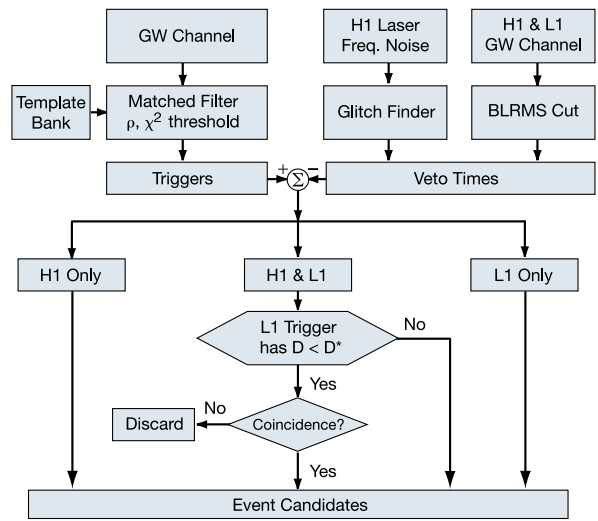}
\end{center}
\caption{\label{f:pipeline}%
The inspiral analysis pipeline used to determine the reported upper
limit.  ``H1 Only'', ``H1 \& L1'', and ``L1 Only'' indicate which
interferometer(s) was/were operating when a trigger was recorded.
This method of recording candidate events even when
coincidence is not available allows a tighter bound to be placed on
the rate of binary neutron star inspirals by providing more
observation time and allowing for the much greater sensitivity of L1
than H1.}
\end{figure}

Our analysis pipeline is summarized in Fig.~\ref{f:pipeline}.  We
follow five steps to produce a list of non-vetoed event candidates which
represent the background due to detector noise (plus any
gravitational-wave signals, if present) during periods of nominal
operation.  (1) Analyze the gravitational-wave channel data from each
detector using matched filtering as described above.  When $\rho > 6.5$ in an
individual detector,  apply
the $\chi^2$ veto to eliminate spurious excitations of the templates.
Store information about the surviving triggers in a database.
(2) Apply the BLRMS cut to reject triggers in periods with unusually high noise, and
apply a veto to eliminate H1 triggers with a clear instrumental origin.
(3) When both interferometers are operating, require coincident
triggers only if the effective distance measured by the L1 detector is
closer than a cut-off distance $D^\ast$. (The selection of 
$D^\ast$ and the coincidence criteria is described below.)
In this case, the SNR for the event candidate is taken to be the L1
SNR in accordance with the discussion around Eq.~(\ref{e:approxsnr}).
If an L1 trigger has $D_{\mathrm{eff}} > D^\ast$, keep the trigger
regardless of whether it was also detected by H1.
(4) During times when only one interferometer is operating, keep any
trigger that passes the cuts in the second step.   %
(5) Finally,  maximize all surviving triggers over time and over the template
bank.  The timing
resolution of inspiral signals is $\lesssim 1$~ms once
coincidence of template mass parameters in both instruments is
enforced.   When coincidence is unavailable, background noise can
trigger many templates at significantly different times. Since the
impulse response of the matched filter is $\sim 16$~seconds 
[because the template is effectively convolved with the frequency dependent 
weighting $1/S_n(f)$ when computing the SNR in Eq.~\ref{e:xfilter}],
we maximize over all triggers in a $16$~second window and over the
entire template bank to
produce the final list of candidate events.  The post-processing analysis
described by steps (2)--(5) was performed using software in the
package \textsc{lalapps}~\cite{LAL}.

We characterized our analysis pipeline using a Monte Carlo method in
which we re-analyzed the data with simulated inspiral signals injected
into the time series.  The re-analysis used exactly the same pipeline
as the original analysis and the simulated signals were drawn from the
population described in Sec.~\ref{s:wavepop}.   The \emph{efficiency}
of the pipeline is the fraction of this population that could be
detected.   To avoid statistical bias, we used only
the playground data set described in Sec.~\ref{s:vetoes} when deciding
aspects of the pipeline.  

The coincident event selection criteria in
step (3) were tuned by studying the fractional loss of efficiency of
the pipeline.   
A trigger from H1 was considered coincident with a trigger from L1 if
the recorded coalescence times were within a time window $\Delta t^* =
0.011 \mbox{ s}$.  This accounts for the light travel time between
the two sites (which is $0.010\mbox{ s}$) plus statistical and
systematic errors in the individual measurements of coalescence time.
The gross
frequency evolution of an inspiral chirp signal is controlled by the
\emph{chirp mass} ${\mathcal{M}}=m_1^{3/5}m_2^{3/5}(m_1+m_2)^{-1/5}$.
The difference of chirp mass $\Delta{\mathcal{M}} = \mathcal{M}_{\text{L1}}
- \mathcal{M}_{\text{H1}}$ for a pair of coincident (in time) triggers was
required to satisfy $|\Delta{\mathcal{M}}|/\mathcal{M}_{\text{L1}} < 10^{-2}$
leading to $\sim 1\%$ fractional loss of efficiency for the playground data.      
Finally,  we chose
$D^\ast = \dstar$~kpc,  producing $\sim 10\%$ fractional loss of efficiency 
for the playground data,
in order to have a reasonable 
chance of detection in coincidence between the two sites.  

\section{Results from S1 data}\label{s:results}

The non-playground data was analyzed using the pipeline described above.
After the division of the data into 256-second blocks, the rejection of blocks
without reliable calibration, the additional loss of 16 seconds from the
beginning of the first block and the end of the last block of a science-mode epoch, and the times during
which a veto was active were discarded, a total of \totaltime\ hours of non-playground
data remained: 58 hours when both L1 and H1 were operating, 76 hours
when only L1 was operating, and 102 hours when only H1 was operating.

\subsection{Triggers and Event candidates}
\label{ss:details}

The triggers from each interferometer satisfy $\rho_{\text{coherent}} > 6.5$ and the
$\chi^2$ veto defined in Eq.~(\ref{e:chisq-threshold}).  There were $\sim
2\times10^6$ triggers from each detector before applying vetoes,
checking for coincidence, and maximizing over templates and time with
a 16~second window.  
The numbers of event candidates from each part of our
pipeline with $\rho_{\text{coherent}} > 8.0$ in the S1 data are 
summarized in Table~\ref{t:results}.\footnote{\label{tablenote}
Since our pipeline with  $\rho_{\text{coherent}} > 6.5$ identifies a high number 
of candidate events (close to the maximum number possible for our pipeline choices),  
we show only candidate events with $\rho_{\text{coherent}} > 8.0$
in Tables~\ref{t:results} and \ref{t:injections}.}

\begin{table}
\caption{\label{t:results}
Number of event candidates with $\rho_{\text{coherent}} > 8.0$ found via 
each of the pipeline paths shown in
Fig.~\ref{f:pipeline}.  The first two lines represent event candidates
found while both interferometers were operating.  No coincident events were
detected in both interferometers; however, there were many event
candidates found in L1 with effective distances
$D_{\mathrm{eff}}>\dstar\textrm{ kpc}$, which would not be detectable
in H1 and thus are kept as event candidates.  The last two lines
represent event candidates found while only one interferometer was
operating.
}\vspace*{0.1in}
\newdimen\digitwidth\setbox0=\hbox{$0$}\digitwidth=\wd0\catcode`?=\active
\def?{\kern\digitwidth}%
\begin{ruledtabular}
\begin{tabular}{llcc}
Operating & Detected in & Number & Max SNR \\\hline
L1 and H1 & L1 ($D_{\text{eff}}<\dstar$ kpc) and H1 & ??0 & -- \\
L1 and H1 & L1 ($D_{\text{eff}}>\dstar$ kpc) & 418 & 15.6 \\
L1 only & L1 & 786 & 15.9 \\
H1 only & H1 & 274 & 12.0
\end{tabular}
\end{ruledtabular}
\end{table}

No event candidates were found in coincidence by both detectors.  If
there had been one or more coincident event candidates,
the background rate of accidental coincidences could have been
determined from the data by counting coincidences after shifting the
H1 trigger times relative to the L1 trigger times by an amount greater
than the light travel time between the sites.    In fact, in the S1
data, there were no triggers whatsoever in L1 which were close enough
($D_{\mathrm{eff}} < \dstar \textrm{ kpc}$) to have been seen in H1
with $\rho_{\mathrm{H1}}>6.5$.

For comparison, Table~\ref{t:injections}$^{\,\textrm{\footnotesize\ref{tablenote}}}$ 
shows the number of events
identified with $\rho_{\text{coherent}} > 8.0$ by the same analysis pipeline 
upon processing the output of the
Monte-Carlo simulation described in Sec.~\ref{s:playground}.  A total of
5071 simulated signals were overlaid on the S1 data, of which 619 were
found in coincidence, demonstrating that the pipeline could correctly
identify coincident event candidates within $\dstar \textrm{ kpc}$.  
Note that the counts of event candidates in the
other three paths of Table~\ref{t:injections} include those in the 
underlying data, not associated with an injected signal.

\begin{table}[h]
\caption{\label{t:injections}
Results from the Monte-Carlo simulation given for comparison with the
equivalent results of the search.   Note that 619 simulated events
were detected in coincidence,  demonstrating that the pipeline was
indeed capable of identifying coincident event candidates.
}\vspace*{0.1in}
\newdimen\digitwidth\setbox0=\hbox{$0$}\digitwidth=\wd0\catcode`?=\active
\def?{\kern\digitwidth}%
\begin{ruledtabular}
\begin{tabular}{llcc}
Operating & Detected in & Number & Max SNR \\\hline
L1 and H1 & L1 ($D_{\text{eff}}<\dstar$ kpc) and H1 & ?619 & 634.4 \\
L1 and H1 & L1 ($D_{\text{eff}}>\dstar$ kpc) & ?773 & ?46.5 \\
L1 only & L1 & 2052 & 460.2 \\
H1 only & H1 & 1623 & 221.9 
\end{tabular}
\end{ruledtabular}
\end{table}

\begin{table*}
\caption{\label{t:loudest}
The five candidates with the
largest SNR which remain at the end of the pipeline.
This table indicates the time they registered in the detectors,  the
SNR,  the value of $\chi^2$ per degree of freedom,
the effective distance to an astrophysical event with the same
parameters, and the binary component masses of the best matching template.
}\vspace*{0.1in}
\newdimen\digitwidth\setbox0=\hbox{$0$}\digitwidth=\wd0\catcode`?=\active
\def?{\kern\digitwidth}%
\begin{ruledtabular}
\begin{tabular}{cccllccccc}
Date & UTC & GPS Time & Operating & Detected in & SNR & $\chi^2/\text{d.o.f.}$
& $D_{\mathrm{eff}}$ (kpc) & $m_1$ ($M_\odot$) 
& $m_2$ ($M_\odot$)\\
\hline
2002/09/02 & 00:38:33.557 & 714962326.557 & L1 only & L1 
& 15.9 & 4.3 & ?95.0 & 1.31 & 1.07 \\
2002/09/08 & 12:31:38.282 & 715523511.282 & L1 and H1 & L1 ($D_{\text{eff}}>\dstar$ kpc)
& 15.6 & 4.1 & ?68.4 & 1.95 & 0.92 \\
2002/08/25 & 13:33:31.000 & 714317624.000 & L1 only & L1 
& 15.3 & 4.9 & 100.7 & 3.28 & 1.16 \\
2002/08/25 & 13:29:24.250 & 714317377.250 & L1 only & L1 
& 14.9 & 4.6 & ?88.7 & 1.99 & 1.99 \\
2002/09/02 & 13:06:56.731 & 715007229.731 & L1 only & L1 
& 13.7 & 2.2 & ?96.3 & 1.38 & 1.38 
\end{tabular}
\end{ruledtabular}
\end{table*}

\begin{figure*}
\begin{center}
\begin{tabular}{cc}
\includegraphics[width=0.45\linewidth]{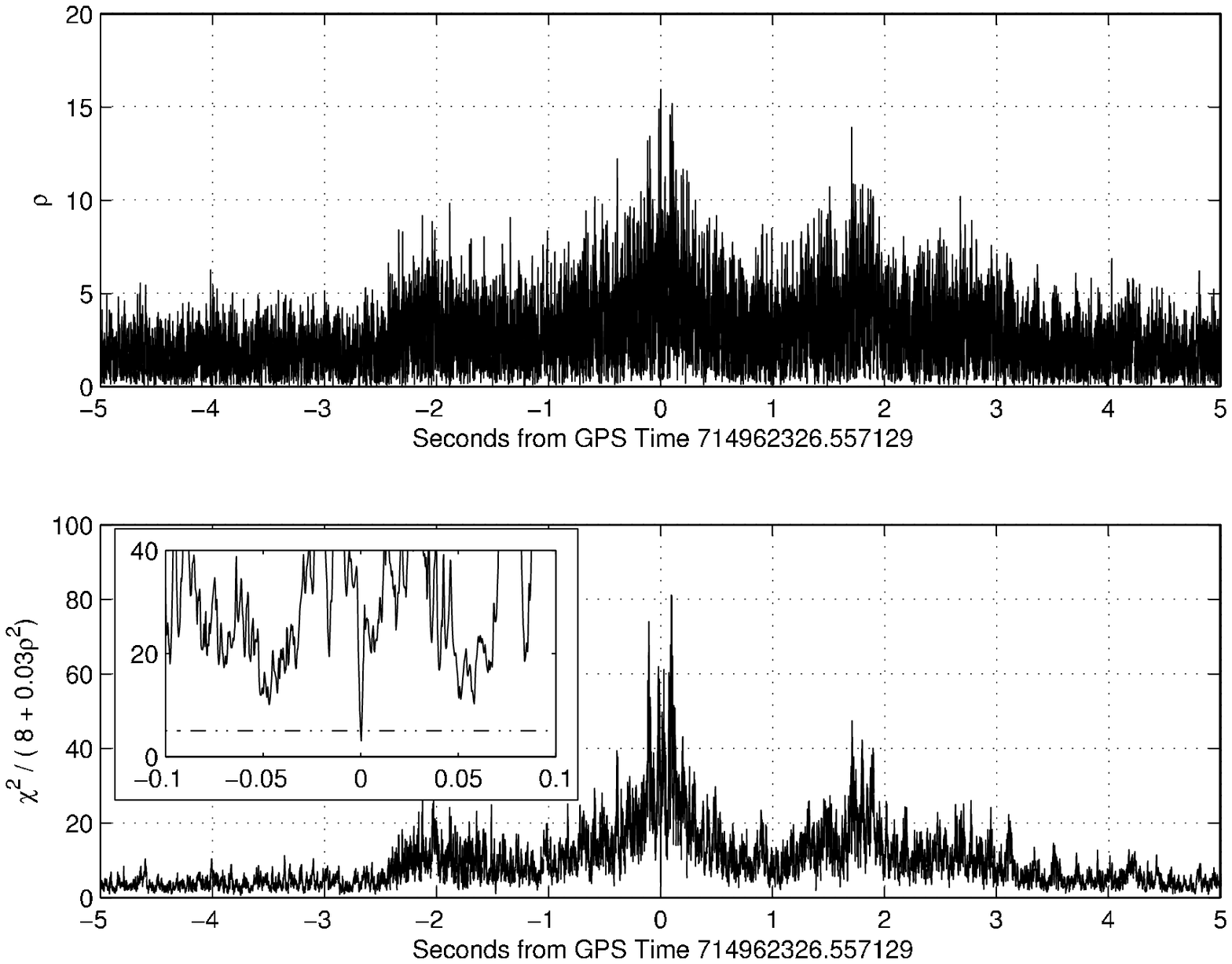} &
\includegraphics[width=0.45\linewidth]{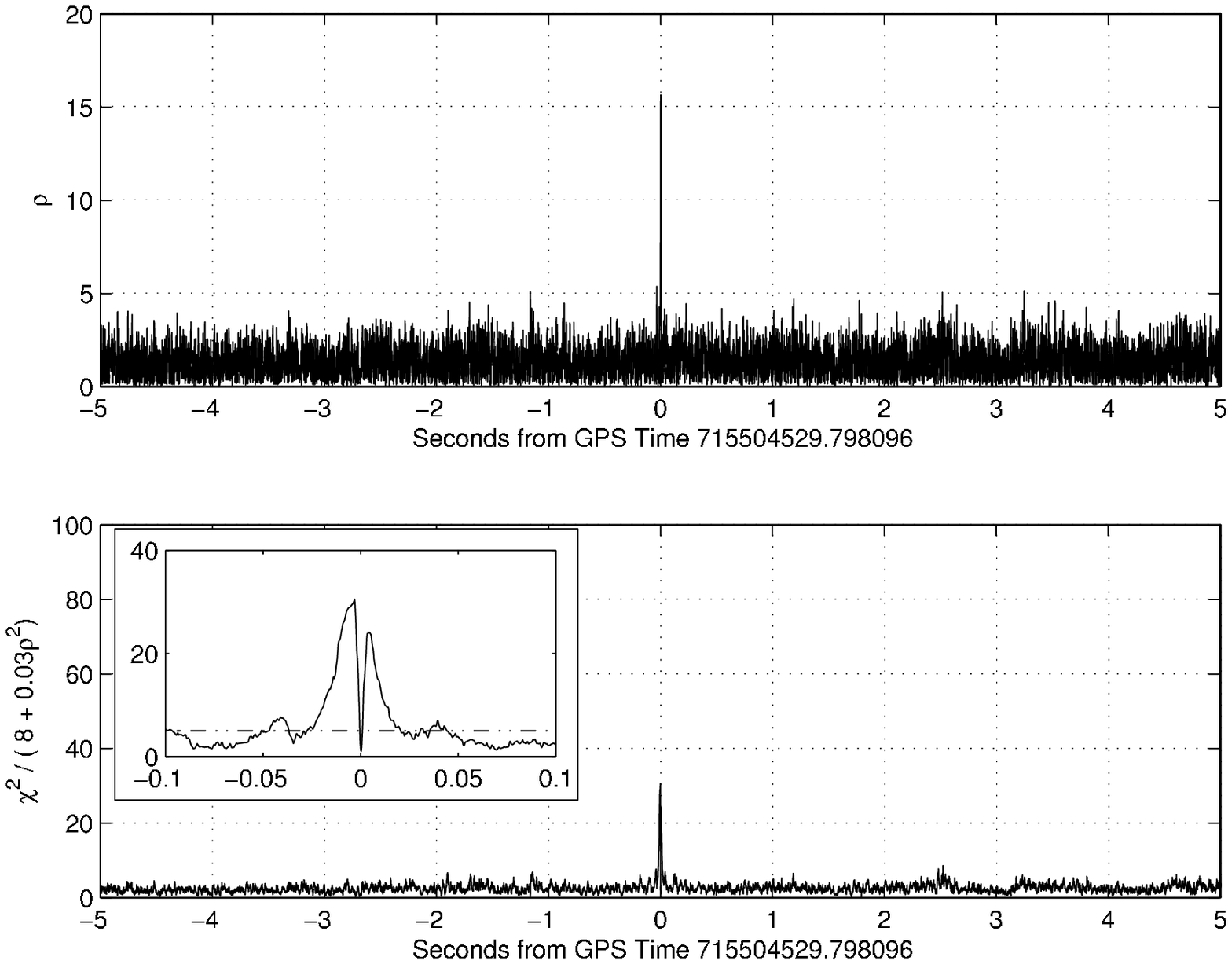}\\
\includegraphics[width=0.45\linewidth]{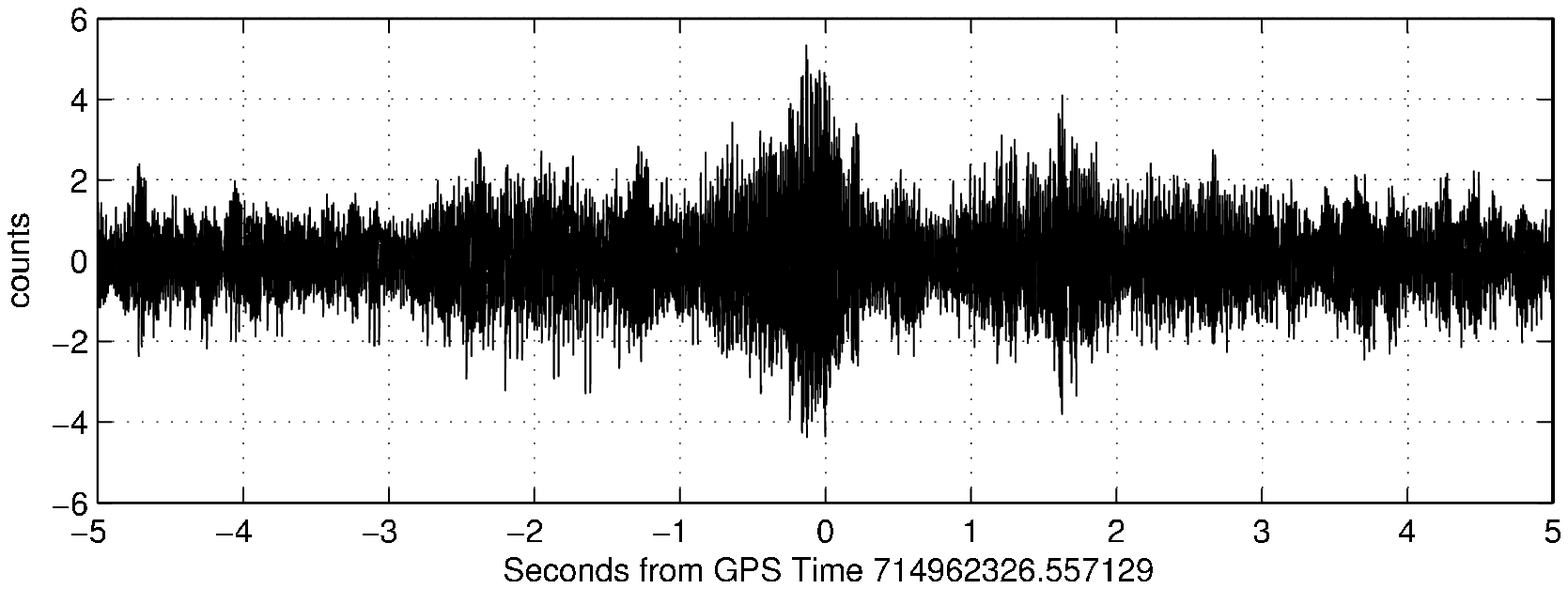} &
\includegraphics[width=0.45\linewidth]{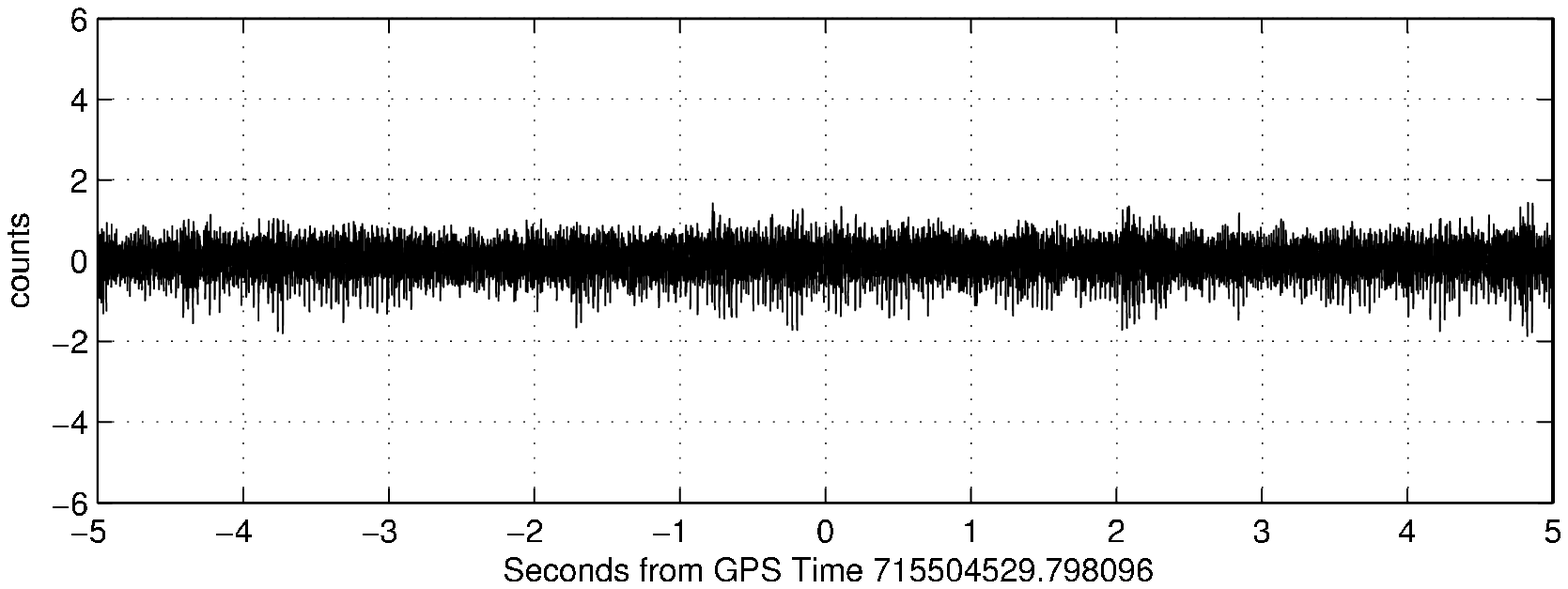}
\end{tabular}
\end{center}
\caption{\label{f:snrplot}
Left Panels: The largest SNR candidate event seen during our search of the
LIGO data.   This candidate event occurred at a time when only L1 was
in stable operation.  The top panel shows the signal-to-noise
time series, $\rho(t)$.  Notice that $\rho(t) > 6.5$ many times in a  $\sim
5$ second interval around the candidate event.   The center
panel shows $\chi^2/ (p+ 0.03 \rho^2)$ as a function of time;  notice $\chi^2 / 
(p+ 0.03 \rho^2) > 5$ for $\sim 5$ seconds around the candidate
event,  but drops below this threshold right at the time of maximum
$\rho$.  The inset shows this more clearly for $\pm 0.1$ second
around the event where the threshold is indicated by a dot-dashed
horizontal line.  The bottom panel shows the time series for this
candidate event after applying a high-pass filter with a knee
frequency of 200~Hz.  Notice
the bursting behavior which does not look like an inspiral chirp
signal.  \break
Right Panels: A simulated injection into the L1 data.  This example was chosen for
comparison with the largest SNR event shown in the left panels since it
similar in mass parameters, detected signal to noise and $\chi^2$.   The
instrument was behaving well at the time around the simulated injection.
The top panel shows that $\rho(t) < 6.5$ except in close proximity to the
signal detection time.   The center panel shows $\chi^2/
(p+ 0.03 \rho^2)$ as a function of time.  Notice that it is much closer
to threshold at all times around the simulated injection; this
contrasts dramatically with the case of the candidate event shown in
the left panels.   The inset shows this more clearly for $\pm
0.1$ seconds around the injection.  The bottom panel shows the
time series for this simulated injection after applying a high-pass
filter with a knee frequency of 200~Hz.  The inspiral chirp signal
is not visible in the noisy detector output.}
\end{figure*}

The ten event candidates with the largest SNR in the pipeline were all
detected by L1 and had SNR between 12
and 16 and 
$\chi^2$ per degree of freedom between 2.2 and 4.9. Details of the
five largest events are given in Table~\ref{t:loudest}.   Four of
these events have $\chi^2$ values close to the threshold in
Eq.~(\ref{e:chisq-threshold}); the exception is the candidate which occurred 
at 13:06:56.731 UTC on 2002/09/02.
Figure~\ref{f:snrplot} (left panels) shows the signal-to-noise and $\chi^2$ time
series for the candidate with the largest SNR, which occurred at 
00:38:33.557 UTC on 2002/09/02.
A simulated inspiral signal with comparable SNR is shown
in Fig.~\ref{f:snrplot} (right panels) to demonstrate the qualitative
differences in the time series.   Unlike the simulated signal,
the SNR of the event
candidate is consistently high across the duration of the
event, with the value of the $\chi^2$ veto varying significantly and
dropping below the threshold right at the time of maximum SNR.

Further scrutiny of the five largest SNR events revealed some instrumental
problems. The event at 00:38:33.557 UTC on 2002/09/02
coincides in time with saturation of the photo-diode at the
antisymmetric port. This saturation, which started a second before the
recorded coalescence time for the candidate event and lasted several 
seconds, was likely due to an instrumental misalignment. The misalignment 
is indicated by a five-fold increase in the power at the
dark port of the interferometer, starting three seconds before the
coalescence time and lasting six seconds.
This event would have been vetoed by the auxiliary-channel veto condition we
considered for L1 but decided not to use (as discussed in Sec.~\ref{ss:vetoes}).
The event recorded at 13:06:56.731 UTC on 2002/09/02
occurred when 
the interferometer was kept functioning during the most severe seismic
conditions for S1 data.  Another event candidate, with SNR $13.0$, occurred
just 98 seconds later. The interferometer was rarely locked with
seismic noise this high, and was probably experiencing up-conversion of
low-frequency seismic noise into the gravitational-wave band through
coupling with mechanical resonances and power line harmonics.

Event candidates detected in just one interferometer cannot be taken
to be real gravitational wave inspirals with any confidence, since we
do not understand the distribution of background.  However, we
can still place an \emph{upper limit} on the rate of inspirals.
Despite being able to find {\it a posteriori} reasons
to justify eliminating some of the largest SNR event candidates as instrumental
effects, we chose to keep them as event candidates
for purposes of calculating the upper limit.

\subsection{Upper Limit Analysis}
\label{ss:statistics}

To determine an upper limit on the rate of binary neutron star
inspirals,  we compare the observed distribution of events as a
function of $\rho_{\text{coherent}}$ to the expected background plus
the population of interest.      
The comparison is made based on criteria established in advance of the
analysis.  Typically, one might choose an SNR threshold $\rho^\ast$
based on the rate and distribution of background events and compare
the number of observed events with $\rho>\rho^\ast$ to the expected background.
Unfortunately,  we have no model for the background events in each of the
interferometers;   this is problematic because we chose to include
event candidates found in only one interferometer to increase the
visible distance and observation time.  Rather than
choosing a fixed value for $\rho^\ast$, we adopt an approach in which
$\rho^\ast$ is determined by the data.  Specifically, we set
$\rho^\ast$ equal to the largest SNR observed in the data and
calculate the efficiency of the pipeline accordingly.  
Since no events
are observed with $\rho>\rho^\ast$, we calculate
an upper limit on the event rate for the modeled population assuming
the probability of a background event above this SNR is negligibly
small.   This approach has the advantage of dealing with the lack of a
model for the background events in a controlled manner. 

If the population of sources produces
Poisson-distributed events with a rate $\mathcal{R}$,  the efficiency
$\epsilon(\rho^\ast)$ is also the probability that any given binary neutron star
inspiral in the target population would have SNR greater than $\rho^\ast$.
Then the probability of observing an inspiral signal with $\rho >
\rho^\ast$, given some rate $\mathcal{R}$ and some observation time $T$, is
\begin{equation}
  P(\rho>\rho^\ast;{\mathcal{R}}) = 1 - e^{-{\mathcal{R}}T\epsilon(\rho^\ast)}.
\end{equation}
A frequentist upper limit with $90\%$ confidence on the value of $\mathcal{R}$ is
determined by solving $P(\rho>\rho_{\text{max}};{\mathcal{R}}_{90\%})=0.9$ for 
${\mathcal{R}}_{90\%}$ 
where $\rho_{\text{max}}$ is the largest SNR event observed in the S1 data. 
The result can be written in closed form as
\begin{equation}
{\mathcal{R}}_{90\%}=\frac{2.303}{ T \epsilon_{\text{max}} }
\end{equation}
where $\epsilon_{\text{max}}=\epsilon(\rho_{\text{max}})$ and $T$ is
the observation time.
For ${\mathcal{R}}>{\mathcal{R}}_{90\%}$, there is more than $90\%$
probability that at least one true inspiral event would be observed
with SNR greater than $\rho_{\text{max}}$.
This limit is conservative since the non-zero probability that a
background event could have SNR greater than $\rho_{\text{max}}$ has
been neglected. 

\begin{figure}
\begin{center}
\hspace*{-0.2in}\includegraphics[width=\linewidth]{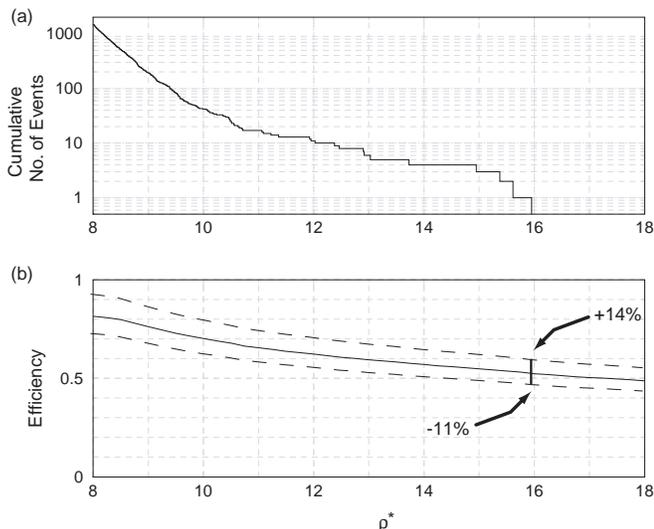}
\end{center}
\vspace*{-0.3in}
\caption{\label{f:events}%
Panel (a) shows the number of events in the data with $\text{SNR} > \rho^\ast$
as a function of $\rho^\ast$.   The largest event has $\text{SNR} =
15.9$.   Panel (b) shows the detection efficiency $\epsilon(\rho^\ast)$ for 
sources in the target population (Milky Way and Magellanic Clouds) as 
a function of $\rho^\ast$.  The dashed lines indicate boundaries of our
estimated systematic errors on the efficiency.   
}
\end{figure}

It is useful to express the limit as a rate per Milky-Way Equivalent
Galaxy (MWEG) for easy comparison with theoretical predictions and
other observational results. 
The effective number of Milky Way equivalent galaxies to which the
search was sensitive is
\begin{equation}
N_{\text{G}} = \epsilon_{\text{max}} 
\left(\frac{L_{\text{pop}}}
{L_{\text{G}}} \right)
\end{equation}
where $L_{\text{G}} = 9 \times 10^9 L_\odot$ is the effective
blue-light luminosity of the Milky Way and $L_{\text{pop}}$  is the
effective blue-light luminosity of the population.  
The rate limit can be written as 
\begin{equation}
{\mathcal{R}}_{90\%}={2.303} \times 
\left( \frac{1 \text{ y}}{T} \right)
\left( \frac{1}{N_{\text{G}}} \right)
\text{ y}^{-1}
\text{ MWEG}^{-1} \; .
\end{equation}
During the $T=\checked{\totaltime}\text{ h} = 0.027\text{ y}$ of
data used in our analysis, the largest observed SNR was
$\rho_{\text{max}} = \rhomax$.  The detection efficiency was computed
using a Monte Carlo simulation in which we re-analyzed the data with
simulated inspiral signals, drawn from the population described in
Sec.~\ref{s:wavepop},  injected into the time series. The efficiency
$\epsilon(\rho^\ast)$,  shown in Fig.~\ref{f:events}(b), is the fraction
of the 5071 simulated signals which were detected with $\rho > \rho^\ast$.
The efficiency at $\rho^\ast = \rhomax$ is $\epsilon_{\text{max}} =
\checked{\efficiency}$.  Folding this together with $L_{\text{pop}}
= 1.13 L_{\text{G}}$, the nominal value of $N_{\text{G}}$ is $\nmweg$;
however, this is subject to some uncertainties, to be discussed
in the next section.  As a function of the true value of
$N_{\text{G}}$, the rate limit is
\begin{equation}
\label{eq:limitng}
{\mathcal{R}}_{90\%}=\rpermweg \left( \frac{\nmweg}{N_{\text{G}}}
\right) \text{ y}^{-1}
\text{ MWEG}^{-1} \; .
\end{equation}

It is interesting to compare our result with a direct estimate based
on average sensitivity of the instruments (as shown in
Fig.~\ref{f:sensitivity}),  properties of the population, and the observation
times used in this analysis.   At SNR $15.9$,  L1 was sensitive to
$80\%$ of the sources and H1 was sensitive to $35\%$ of sources in our
model population.   Out of 236 hours,  L1 was the best detector for
134 hours and H1 for 102 hours.  The expected efficiency is then
\begin{equation}
   \epsilon(15.9) =
     (102 \times 0.35 + 134 \times 0.80)/236 = 0.6 \; .
\end{equation}
The measured efficiency is $\epsilon(15.9) = \efficiency$,  but
the $\chi^2$ veto and coincidence requirements both introduce some
loss;  the expectation based on playground data was $\approx 0.06
\times 58 / 236 = 0.015$
decrease in efficiency from coincidence and a loss of about
$\approx 0.06$ from the $\chi^2$.   The actual loss from coincidence is
$\approx 0.02$ as measured on the full data
set.  Consequently,  the measured efficiency and hence the upper limit agree 
well with expectations.

\section{Error analysis}
\label{s:errors}

The interpretation of this search for gravitational waves from binary
neutron star inspiral suffers from a number of systematic effects
which could modify the upper limit.    We classify
these effects into three different types:  (i) uncertainties in the
population model and theoretical expectations about the sources;  (ii)
uncertainties in the instrumental calibration;  (iii) deficiencies of
the analysis pipeline.    Each one can have a direct effect on the
efficiency of the search to detect gravitational waves from the target
population as it exists in nature.    

\subsection{Uncertainties in population model}
\label{ss:popuncertainties}
Uncertainties in the population model used for the Monte-Carlo
simulations may lead to differences between the inferred rate and the
rate in the universe. Since the effective blue-light luminosity
$L_{\text{pop}}$ is normalized to our Galaxy,   variations arise
from the relative contributions of other galaxies in the population.
These contributions depend on the estimated distances to the galaxies,  
estimated reddening,  and corrections
for metallicity (lower values tend to produce higher mass binaries),
among other things.  Since the Magellanic Clouds contribute only $\sim
13\%$ of the blue light luminosity in this analysis,  a conservative
estimate of the uncertainties gives $L_{\text{pop}} = 1.13 \pm
0.06$.   

The spatial distribution of the sources can also introduce significant
uncertainties. Typically, the distances to nearby galaxies are only known
to about 10\% accuracy. Uncertainties in distances to galaxies near
the limit of detector sensitivity are most relevant.  As the
detector sensitivity improves, more galaxies will be in this
category, so it may become a major source of systematic uncertainty. It
is not important for the current analysis,  since the detectors were
sensitive to the majority of sources in the Milky Way and Magellanic
Clouds.  

The effects of spin were ignored both in the population and in the
waveforms used to detect inspiral signals.
Apostolatos~\cite{Apostolatos:1995} has performed the most complete
analysis of the effects of spin on detection of waves from neutron
star inspiral.   His investigations suggest that less than $10\%$ of
all possible spin orientations cause more than $\sim5\%$ reduction in SNR for
binary neutron star systems.   There is insufficient information about
the distribution of binary spin orientations to quantitatively
estimate the systematic effect, but it seems certain that the fraction
of the population with spin configurations which would interfere with
their detection is negligible.

Different models for NS-NS formation can lead to small variations in
the tails of the NS mass distribution~\cite{Belczynski:2002}, but the bulk of the
distribution always remains strongly peaked around observed NS
masses~\cite{Thorsett:1998uc}. Since the detection
efficiency depends most sensitively on the bulk properties of the mass
distribution,  the expected variations are negligible compared to other
systematic effects discussed in this section.

\subsection{Uncertainties in the instrumental response}
\label{ss:respuncertainties}

The instrument response $R(f)$ was constructed for every minute of
data during S1 from a reference sensing function $C(f)$, a reference
open loop gain function $G(f)$, and a parameter $\alpha(t)$
representing varying optical gain~\cite{Gonzalez:2002}. The parameter
$\alpha$ was reconstructed using the observed amplitudes of the
calibration lines described in Section~\ref{s:instrument}.
If an inspiral signal is present in the data, systematic errors in the
calibration can cause a mismatch between the template and the
signal.
For simulated injections,  the SNR
differs from the SNR that would be
recorded for a signal from a real inspiral event at the same distance
as the injection.   The effect is linear in amplitude errors
causing either an upward or downward shift in SNR,
but quadratic in phase errors causing an over-estimation of
sensitivity.   This effect is captured by shifting the efficiency curve in
Fig.~\ref{f:events} horizontally by the appropriate amount.

A careful evaluation of uncertainties in the S1 calibration~\cite{Gonzalez:2002}
has shown that amplitude errors are
primarily due to statistical fluctuations in the measurement
procedure, while phase errors are mostly systematic and are greater at
higher frequencies.  Combining statistical and systematic errors in
quadrature, the amplitude errors lead to $\sim 18\%$ errors in SNR in
L1 and $\sim 8\%$ errors in H1.  The phase errors lead to
overestimation of the SNR by $\sim 2\%$ in L1 and $\sim 4\%$ in H1.
Combining amplitude and phase errors in quadrature and taking the 
larger L1 values as representative, we find $\sim 18\%$ errors in
SNR of Monte-Carlo injections which translates to fractional errors in
efficiency $\sim \plusminus{14\%}{10\%}$,  i.e. $\epsilon_{\text{max}}
= \efficiency^{+0.07}_{-0.05}$.

To verify the data analysis methods, a few special studies were done
in which simulated inspiral waveforms were injected into the
interferometer hardware using the mirror actuators.
We then used the analysis
pipeline described above to recover the known mass and distance
parameters of the injected signal.  A side benefit of these injections
is to build confidence in our understanding of calibration uncertainties. 
In order to simplify the analysis pipeline, the template bank was
reduced to a single template, a $1.4,1.4\,M_\odot$ or a
$4.0,1.4\,M_\odot$ inspiral, corresponding to the mass parameters of
the injected signal.   Unfortunately,  the calibration signal was
turned off during the injections,  so we defined a
set of possible response functions for this range, and studied the
variation in the detected inspiral signal.  This was possible because
the parameter $\alpha$ has only a limited physical range. We found that
the variation in the reconstructed signal to noise and effective distance
was in agreement with our expectations.
Since the parameter $\alpha$ has a known dependence on the interferometer
alignment we were able to use auxiliary channel information to estimate
its value during the injections.
For this value the detected coalescence time of the chirp was the same 
as the injected
time to within $1/4096$ seconds, i.e. one sample of filtered data, and
the reconstructed distance and the injected distance agreed to within
12\%, which is consistent with the errors quoted above. 

\subsection{Uncertainties in the analysis pipeline}
\label{ss:pipuncertainties}
Since we use matched filtering to search for gravitational waves from
inspiralling binaries, differences between the theoretical and the
real waveforms could also adversely effect the results.  These effects
have been studied in great detail for binary neutron star
systems~\cite{Apostolatos:1995, Droz:1997, Droz:1998}.  The results indicate
$\sim 10\%$ loss of SNR due to inaccurate modelling of the waveforms
for binaries in the mass range of interest.  This feeds into our
result through our measurement of the efficiency.  We may be
over-estimating our sensitivity to real binary inspiral signals;
this would shift all points on the efficiency curve in Fig.~\ref{f:events} 
to the left by $\sim 10\%$.   This corresponds to fractional errors 
$\sim \plusminus{0\%}{5\%}$ in efficiency,  i.e. $\epsilon_{\text{max}}
= 0.53^{+0.0}_{-0.03}$.     

The effects of discreteness of the template placement,  errors in the
estimates of the power spectral density $S_n(f)$ used in the matched
filter in Eq.~(\ref{e:xfilter}),  and trends in the instrumental noise
are all accounted for by the Monte-Carlo simulation.

\subsection{Combined uncertainties on $N_{\text{G}}$ and the rate}
\label{ss:combuncertainties}
The efficiency incurs fractional errors $\sim \plusminus{14\%}{10\%}$ 
from calibration uncertainties
(Sec.~\ref{ss:respuncertainties}) and  $\sim \plusminus{0\%}{5\%}$ from
inaccurate knowledge of the inspiral waveforms
(Sec.~\ref{ss:pipuncertainties}).  Combining these in quadrature
yields total errors $\sim
\plusminus{14\%}{11\%}$ in the efficiency $\epsilon_{\text{max}}$. 
Adding these (not in quadrature) to the $\pm 5\%$ error for $L_{\text{pop}}$
(Sec~\ref{ss:popuncertainties}) yields
\begin{equation}
N_{\text{G}} = \nmweg^{+0.12}_{-0.10} \; .
\end{equation}
To be conservative, we assume the downward
excursion $N_{\text{G}} = \nmweg - 0.10 = 0.50$ when using
Eq.~(\ref{eq:limitng}) to derive an observational
upper limit on the rate of binary neutron star coalescence:
\begin{equation}
{\mathcal{R}} < \ratepermweg \text{ y}^{-1} \text{ MWEG}^{-1} \; .
\end{equation}

\section{Conclusions and future directions}\label{s:future}
The first search for gravitational-wave signals from coalescing
neutron stars in LIGO science data yielded no coincident event
candidates.   An observational upper limit $\ratepermweg \text{
y}^{-1} \text{ MWEG}^{-1}$ on the rate of neutron star inspirals
was derived.  This limit is better than previous direct limits by a
factor of 26~\cite{Allen:1999yt,Tagoshi:2000bz}.

Over the next few years, the sensitivity of the LIGO interferometers
will be dramatically improved, to the point where inspirals of double
neutron stars ($ 2 \times 1.4 M_\odot$) are expected to be detectable
within an equivalent volume $\approx (4 \pi /3) \times (21  \textrm{
Mpc})^3$~\cite{Finn:2001js}.   Due to the non-uniform response of the
detectors,  this implies that a neutron star inspiral could be detected out to
a maximum distance $\approx 46 \textrm{ Mpc}$ if the binary is located
directly above or below the detectors and orbits in the plane of the
sky.   
The rate of coalescence of extra-galactic neutron
star binaries is thought to be proportional to the rate of massive
star formation which is, in turn, proportional to the blue light
luminosity. (See,  for example,  Ref.~\cite{Kalogera:2000dz}.)  Using
current galaxy catalogs,  it is estimated that $N_{\text{G}} \approx 465$ MWEG
will be detectable by LIGO (using the three detectors combined to
produce a network SNR $> 8$)~\cite{Nutzman:2003}.  If the coalescence rate 
of binary systems (in which each component has a mass in the range 
$1$--$3$~$M_\odot$) were as high as $\sim 5 \times 10^{-4}
\text{ y}^{-1} \text{ MWEG}^{-1}$~\cite{Kalogera:2000dz,Kim:2002uw},
then the event rate detectable by
LIGO would be $N_{\text{G}}$ times higher providing up to 1/4 events
per year.  
In lieu of a detection, an upper
limit within the range of astrophysical expectations will
constrain the binary neutron star population models, and especially the
population of electromagnetically-undetectable pulsars at the faint end of
their luminosity function~\cite{Kalogera:2000dz,Kim:2002uw}.

The methods used, and experience gained, on the 17-day S1 data set
will be enhanced and used in future searches for gravitational waves
from coalescing compact binaries with LIGO data. We can expect
improvements in the upper limits obtained with detectors of better
sensitivity, but we can also draw lessons on the methods used from this first 
experience.  For example, we expect to reduce the maximum SNR of
non-gravitational wave signals by making better use of the knowledge
of the instrument status to find more efficient veto criteria. 
\textnote{DHS}{
In our next search, we will require coincidence from candidates from
the two observatories to establish an event.} This will allow us to
to measure a background rate of accidental coincident events, using
techniques to find lower SNR triggers as needed in the least sensitive
instrument (if there continue to be significant differences in
sensitivities). Eventually, we would like to use coherent methods with
all the different detectors in operation. Even though the errors in
the upper limits obtained in this article do not compromise their
significance, the same errors would affect more seriously the
parameter identification of a detection, so we hope to improve on all
aspects contributing to statistical and systematic errors.

Future searches will also target neutron-star/black-hole and
black-hole/black-hole binaries which produce more energy in
gravitational waves
and will be visible within a much greater volume of the
Universe.  It is possible that several black-hole binaries could be
detected by the initial LIGO
interferometers~\cite{Belczynski:2002,PortegiesZwart:2000}, but there is
considerable uncertainty in this event rate.  An observational upper
limit would constrain population models and yield information about
the formation mechanisms of black-hole binaries.  The challenge of
setting an upper limit on higher-mass binary systems is formidable:
massive binary systems (black-hole/black-hole) will exhibit highly
relativistic effects (beyond the realm of the standard post-Newtonian
approximation) within the sensitivity band of the
instruments~\cite{Damour:2000zb,BuonannoChenVallisneri:2003a}, whereas
spin-orbit and spin-spin coupling in precessing binaries will be
extremely important in intermediate-mass systems of low mass ratio
(neutron-star/black-hole)
\cite{Apostolatos:1995,Apostolatos:1996rg,Apostolatos:1994,GrandclementKalogeraVecchio:2003,Kalogera:2000}.
These effects will greatly expand the parameter space that needs to be
searched, and will require the construction of both
accurate~\cite{Damour:2000zb} and computationally efficient waveforms.
Efforts are already under way to construct \emph{detection template
families} \cite{BuonannoChenVallisneri:2003a,
BuonannoChenVallisneri:2003b,GrandclementKalogera:2003} in our search
codes. The goal with these detection template families is to
efficiently mimic all the known analytical models of black-hole binary
dynamics (such as the standard post-Newtonian
models~\cite{Blanchet:1996pi} and their improved versions, namely,
P-approximants~\cite{Damour:1998zb} and effective one-body
techniques~\cite{BuonannoDamour:1999,BuonannoDamour:2000,Damour:2000zb,DamourJaranowskiSchaefer:2000,Damour:2001})
and/or the effects of precession on waveforms emitted by binaries with
spinning compact objects.  Despite the challenges, a search for
gravitational waves from black hole binaries is the highest priority
for current research.

Another class of systems is the sub-solar mass ($\mbox{0.2--1}\,M_\odot$)
binary black holes that might form a sizable portion of macroscopic
halo objects (MACHOs)~\cite{Nakamura:1997sm}.  If such objects exist, then many of the
challenges in detecting binaries with stellar-mass are alleviated: the orbits
of these binaries will not be highly relativistic while the gravitational waves
are emitted in the LIGO sensitivity band, and the spin effects can be handled
easily.
On the other hand, the smaller amplitude of the gravitational waves
emitted by these sources limits the distance to which they can be
seen.

\acknowledgments
The authors gratefully acknowledge the support of the United States
National Science Foundation for the construction and operation of the
LIGO Laboratory and the Particle Physics and Astronomy Research
Council of the United Kingdom, the Max-Planck-Society and the State of
Niedersachsen/Germany for support of the construction and operation of
the GEO600 detector. The authors also gratefully acknowledge the
support of the research by these agencies and by the Australian
Research Council, the Natural Sciences and Engineering Research
Council of Canada, the Council of Scientific and Industrial Research
of India, the Department of Science and Technology of India, the
Spanish Ministerio de Ciencia y Tecnologia, the John Simon Guggenheim
Foundation, the David and Lucile Packard Foundation, the Research
Corporation, and the Alfred P. Sloan Foundation. 
%

\end{document}